\documentclass[twocolumn,showpacs,preprintnumbers,amsmath,amssymb]{revtex4}

\newcommand{\Hi}{\mathcal{H}}

\newcommand{\lv}{\left \vert}
\newcommand{\rv}{\right \vert}
\newcommand{\la}{\left \langle}
\newcommand{\ra}{\right \rangle}
\newcommand{\ket}[1]{\lv #1 \ra}
\newcommand{\bra}[1]{\la #1 \rv}

\begin{document}
\title{Existence of incomparable pure bipartite states in infinite dimensional systems}
\author{Masaki Owari$^{1,3}$, Keiji Matsumoto$^{2,3}$ and Mio
Murao$^{1,4}$}
\address{
$^1${\it Department of Physics, The University of Tokyo, Tokyo 113-0033, Japan}\\
$^2${\it Quantum Computation Group, The National Institute of Information, Tokyo 101-8430, Japan}\\
$^3${\it Imai Quantum Computation and Information Project, JST, Tokyo 113-0033, Japan}\\
$^4${\it PRESTO, JST, Kawaguchi, Saitama 332-0012, Japan} }
\date{\today}

\begin{abstract}
Based on set theoretic ordering properties, a general formulation
for constructing a pair of convertibility monotones, which are
generalizations of distillable entanglement and entanglement cost,
is presented.  The new pair of monotones do not always coincide
for pure bipartite infinite dimensional states under SLOCC
(stochastic local operations and classical communications),
demonstrating the existence of SLOCC incomparable pure bipartite
states, a new property of entanglement in infinite dimensional
systems, with no counterpart in the corresponding finite
dimensional systems.
\end{abstract}


\maketitle

%%%%%%%%%%%%introduction%%%%%%%%%%%%%%%%%%%%%%%%%%%%%%%%%%%%%%%%

Entanglement, or nonlocal quantum correlation, is regarded as the
key resource which allows many quantum information processing
schemes to out-perform their classical counterparts. Better
understanding of entanglement is essential for the field of
quantum information.  Entanglement can be classified by
``plasticity'' under local operations, such as LOCC (deterministic
local operations and classical communications), SLOCC (stochastic
LOCC, nondeterministic LOCC) and PPT
(positive-partial-transpose-preserving) operations \cite{ppt}.
Therefore the convertibility properties of two different entangled
states (in a single copy or multi-copy situation) under local
operations are important for the qualitative and quantitative
understanding of entanglement.

For finite dimensional bipartite systems, we now have a better
understanding of LOCC and SLOCC convertibility based on intensive
work in recent years.  For example, the condition for
convertibility of two pure entangled states in the single copy
situation is given by Nielsen's majorization theorem
\cite{nielsen} for LOCC, and is given by Vidal's theorem
\cite{vidal} for SLOCC.  On the other hand, for infinite
dimensional systems (or continuous variable systems), there are
still many open questions on general LOCC and SLOCC
convertibility, although there are important works \cite{gaussian,
gaussian distill}, which have investigated a limited class of
local operations (gaussian operations).

Infinite dimensional systems have been expected to offer high
potential for quantum information processing. One of the
advantages of infinite dimensional systems is the possibility of
implementation using quantum optical systems, as shown by the
successful demonstration of teleportation \cite{teleportation}.
Another advantage is the possibility of (yet unknown) new types of
quantum information processing schemes, which do not exist in
finite dimensional systems. If such schemes exists, their
existence should be related to essential properties of quantum
systems, such as entanglement. Thus the discovery of entanglement
properties which are unique to infinite dimensional systems is
very desirable.

In a previous paper \cite{owari-locc}, we proved that Nielsen's
majorization theorem and Vidal's theorem can be generalized to
infinite dimensional states by introducing the concept of
$\epsilon$-convertibility. We showed a new classification of
infinite dimensional entangled states based on SLOCC
convertibility by introducing the rapidity of convergence, which
corresponds to Schmidt rank \cite{vidal} in finite dimensional
systems.  The classifications obtained are natural extensions of
those for finite dimensional systems, thus they did not suggest
the existence of new entanglement properties. In this letter, we
will show that a new entanglement property in infinite dimensional
systems, the existence of SLOCC incomparable pure bipartite
states, by developing a formulation for convertibility based on
set theoretic ordering properties.

%%%%%%%%convertibility and set theory%%%%%%%%%%%%%%%%%%%%%%%%%

We consider ordering sets, ordered by the convertibility of two
general bipartite quantum states (including infinite dimensional
states) under general operations.  For such sets, we can define a
pair of monotones of the given ordering based on set theory. The
definition of these monotones is the core element of our
formulation. If these two monotones coincide, it indicates that
the ordering is a total ordering, namely, at least one of the
states can be converted to another. For the case of finite
dimensional pure states under SLOCC operations, the two monotones
coincide. In contrast, we will show that the two monotones do not
always coincide in the infinite dimensional case. Thus the
ordering can be non-total (partial) and the two states can be
incomparable.

In this letter, we first present the construction of the general
formulation. Then, we apply it to infinite dimensional pure states
under SLOCC operations. The advantage of our formulation is that
it can be applied for many different situations: single-copy or
asymptotic (infinitely many-copy) cases, for finite or infinite
dimensional systems, mixed or pure states, and under LOCC, SLOCC
or PPT operations. It should be noted that the two monotones
become distillable entanglement and concentration \cite{bennett}
for mixed states in the asymptotic situation, therefore, they can
be considered as a generalization of distillable entanglement and
entanglement cost.

In the language of set theory, convertibility of physical states
under some operation can be describe by an order denoted by
``$\rightarrow$''. For a set of physical states $S$, the order
indicating the existence of physical transformation satisfies the
reflective law $a \rightarrow a$ and the transitive law $a
\rightarrow b$ and $b \rightarrow c$ imply $a \rightarrow c$),
where $a, b, c \in S$. This ordering property is pseudo partial
ordering. We consider that two states are in a same equivalence
class, if they transform each other $a \rightarrow b$ and $b
\rightarrow a$. We denote this situation as $a \leftrightarrow b$.
The quotient set of $S$ by the equivalent class $\leftrightarrow$
is denoted as $(S/\leftrightarrow , \rightarrow)$ and represents
the classification based on the given transformation.  For the set
$(S/\leftrightarrow , \rightarrow)$, we can redefine the ordering
$\rightarrow$ which satisfies the additional condition $a
\rightarrow b$ and $b \rightarrow a$ imply $a = b$. This ordering
property is partial ordering.

If the set has the additional property $a \nrightarrow b$ implies
$b \rightarrow a$ ($a \nrightarrow b$ denotes $a \rightarrow b$ is
not true), the set is totally ordered.  Total ordering is a
convenient property to analyze the convertibility of a system,
since there exists a unique measure of ordering for a totally
ordered set. For example, the convertibility of pure state
$\ket{\phi}$ under LOCC transformation in the asymptotic situation
is total ordering. Therefore, there is a unique measure, or a
monotone, given by the von Neumann entropy of entanglement
$E(\ket{\phi})$ in this case \cite{thermodynamics}. On the other
hand, a pair of monotones, like distillable entanglement and
entanglement cost, are useful tools to distinguish the ordering
properties (total ordering or partial ordering) of the system
\cite{morikoshi}.

%%%%%%%%%%%%%%%%%%%ordering-monotone%%%%%%%%%%%%%%%%%%%%%%%%%%%%%%%%

The key idea of our formulation is that we consider a totally
ordered subset $\{ \xi _r \}$ parameterized by a real number $r$
for a pseudo partial ordering set $S$.  Then we can always define
a pair of functions $R^-(\psi)$ and $R^+(\psi)$ for a state $\psi
\in S$, where $R^-(\psi)$ is the supremum of $r$ at which $\psi$
can be transformed to $\xi _r$, and $R^+(\psi)$ is the infimum of
$r$ at which $\Psi$ can be transformed to $\xi _r$. Mathematically
they are expressed as the following: For a pseudo partial ordering
set, if there exists a real parameterized total ordering subset
$\{ \xi _r \} _{r \in A} \subset S$ where $A \subset \mathbb{R}$
such that $r_1 \le r_2$ if and only if $\xi _{r_1} \rightarrow \xi
_{r_2}$, we can define a pair of functions on $S$ to $\overline{A}
\subset \mathbb{R}$ as
\begin{eqnarray}
R^-(\psi) &=& \sup \{ r \in A | \psi \rightarrow \xi _r \}
\label{rminus}
\\
R^+(\psi) &=& \inf \{ r \in A | \xi _r \rightarrow \psi \}
\label{rplus}
\end{eqnarray}
where we define $R^-(\ket{\psi}) = \inf \{ A \}$ for $\{ r\ \in A
| \psi \rightarrow \xi _r \} =\emptyset $, and $R^+(\ket{\psi}) =
\sup \{ A \}$ for $\{ r\ \in A | \xi _r \rightarrow \psi \}
=\emptyset $.

Although there are many ways to define a monotone for a partial
ordering set from a totally ordered subset $\{ \xi _r \} _A$, our
definition of functions $R^-(\psi)$ and $R^+(\psi)$ are preferable
for analyzing entanglement convertibility. We prove by
contradiction that they are the {\it unique monotones} which give
lower and upper bounds of any monotones defined for a given pseudo
partial ordered set $S$. For simplicity we only show the proofs
for the case of $A$ to be an interval of $\mathbb{R}$ in this
letter, but the proofs can easily be extended to an arbitrary
subset of $\mathbb{R}$.

First, we show that $R^-(\psi)$ and $R^+(\psi)$ are monotones
which satisfy $R^-(\psi) \le R^+(\psi)$ for all $\psi \in S$.
Suppose $R^-(\psi) > R^+(\psi)$, then there exists $r_1$ and $r_2$
such that $R^-(\psi) > r_2 > r_1 > R^+(\psi)$. Since $\xi _{r_1}
\rightarrow \psi$ and $\psi \rightarrow \xi _{r_2}$, $\xi _{r_1}
\rightarrow \xi _{r_2}$. This means $\xi _{r_1} \leftrightarrow
\xi _{r_2}$. This contradicts the totality ordering of $\{ \xi _r
\} _A$, therefore we have $R^-(\psi) \le R^+(\psi)$ for all $\psi
\in \epsilon$. Next, suppose $R^-(\phi) > R^-(\psi)$ and $\psi
\rightarrow \phi$, there exists $r \in A$ such that $R^-(\phi) > r
>  R^-(\psi)$.  Then, $\phi \rightarrow \xi _r$. Since $\psi
\rightarrow \phi$, we have $\psi \rightarrow \xi _r$. This
contradicts $R^-(\phi) > r$, thus we have proved $\psi \rightarrow
\phi$ implies $R^-(\psi) \ge R^-(\phi)$. Similarly, we can prove
$\psi \rightarrow \phi$ implies $R^+(\psi) \ge R^+(\phi)$. In
addition, suppose $R^-(\phi) > R^+(\psi)$, then there exists $r
\in A$ such that $R^-(\phi) > r > R^+(\psi)$. Since $\phi
\rightarrow \xi _r$ and $\xi _r \rightarrow \psi$, we have $\phi
\rightarrow \psi$. Thus we have proved $R^+(\phi) < R^-(\psi)$
implies $\psi \rightarrow \phi$.

Second, we show that $R^-(\psi)$ and $R^+(\psi)$ are lower and
upper bounds of monotones, respectively.  Suppose there was
another monotone $R_0 (\psi)$ defined for a given ordering such
that $\psi \rightarrow \phi$ implies $R_0 (\psi) \ge R_0 (\phi)$
satisfying $R_0 (\xi _r) = r$ for all $r \in A$. If $R_0 (\psi) <
R^-(\psi)$, there exists a real number $r \in A$ such that $R_0
(\psi) < r < R^-(\psi)$, then we obtain $\psi \rightarrow \xi _r$.
On the other hands, from $R_0 (\psi) < r$ we have $\psi
\nrightarrow \xi _r$ by the monotonicity of $R_0(\psi)$. This is
also a contradiction.  Similarly we can prove $R_0 (\psi) \le
R^+(\psi)$. Thus $R^- (\psi) \le R_0 (\psi) \le R^+ (\psi)$ {for
all $\psi \in S$.

From the properties of $R^-(\psi)$ and $R^+(\psi)$, we can
immediately derive the following important results:  If the
quotient set $(S/\leftrightarrow , \rightarrow)$ is totally
ordered, namely, $\psi \nrightarrow \phi$ implies $\phi
\rightarrow \psi$ is satisfied, then for all $\psi \in S$, we have
$R^-(\psi) = R^+(\psi)$. On the other hand, if there exists $\psi
\in S$ such that $R^-(\psi) < R^+(\psi)$, then $(S/\leftrightarrow
, \rightarrow)$ is not totaly ordered, and $\psi$ is incomparable
to all $\xi _r$ with $R^-(\psi) < r < R^+(\psi)$.

%%%%%%%%%%%%%%%%%%%%%%%%%%%%%%%%%%%%%%%%%%%%%%%%%%%%%%%%%%%%%%%%%%%%%%

Many important known results of entanglement theory can be
re-derived only from simple ordering properties and the
existence of the real parameterized total ordering subset.  To
demonstrate the power of our formulation, we apply the formulation
to the following four situations of finite ($d$) dimensional
systems; A. LOCC operation for pure states in the single copy
situation: $\{\ket{ \xi _k} \}_{k=1}^d $ is the maximally
entangled state in $k$ dimensional systems \cite{deterministic
concentration}; B. SLOCC operation for pure states \cite{vidal}:
$\{ \ket{\xi _k} \}_{k=1}^d$ is also the maximally entangled state
in $k$ dimensional systems; C. LOCC operations for pure states in
the asymptotic situation: $\{ \ket{\xi _s} \} _{s=0}^{\log _2 d}$
is a subset with $E(\ket{\xi _s}) = s$ \cite{thermodynamics,
uniqueness}; D. LOCC operations for mixed states in the asymptotic
situation: $\{ \ket{\xi _s} \bra{\xi _s} \} _{s=0}^{\infty}$ is a
subset of pure states with $E(\ket{\xi _s}) =s$ \cite{bennett}. We
summarize the representation of $R^-$ and $R^+$ for these four
situations in Table~\ref{table}. In the situations A and D, we see
that the sets are not totally ordered and $R^-$ (distillable
entanglement) and $R^+$ (entanglement cost) are limits of other
monotones \cite{uniqueness}. Furthermore for D, we see that there
is a set of states such that $R^-=0$ but $R^+
> 0$, the bound entangled states \cite{boundentangle}.

%%%%%%%%%%table%%%%%%%%%%%%%%%%%%%%%%%%%%%%%%%%%%%%%%%%%%%%%%%%%%%%
\begin{table}
  \centering
\begin{tabular}{*{5}{|c}}
    \hline
    \  & $R^+$, $R^-$ \\
    \hline
    A& $R^-(\ket{\psi}) = [\log _2 \lambda _1 ], R^+(\ket{\psi})
    = R_d(\ket{\psi})$ \\
    \hline
    B&   $R^-(\ket{\psi}) = R^+ (\ket{\psi}) = R_d(\ket{\psi})$ \\
    \hline C&   $R^-(\ket{\psi})=R^+ (\ket{\psi}) =E_d(\ket{\psi})=
    E_c(\ket{\psi})
    =E(\ket{\psi}) $          \\
    \hline
    D& $R^-(\rho)= E_d(\rho), R^+(\rho)=E_c(\rho)$         \\
    \hline
\end{tabular}
  \caption{The monotones $R^-$ and $R^+$ for
    convertibility of a finite dimensional bipartite state in
    several different situations.  The description of situations
    are: A. LOCC, pure and single copy,
    B. SLOCC, pure and single copy,
    C. LOCC, pure and asymptotic, D. LOCC mixed and asymptotic.
     In this table,
    $R_d(\ket{\psi})$ denotes the Schmidt rank,
    $E(\ket{\psi})$ is the amount of entanglement for pure states,
    $E_d(\rho)$ and
    $E_c(\rho)$ are distillable entanglement and entanglement cost
    respectively. }\label{table}
\end{table}

%%%%%%%%%%%%%%%%%%%% slocc-monotone%%%%%%%%%%%%%%%%%%%%%%%%%%%%%%%%%%%%

Now we concentrate on the investigation of SLOCC convertibility
(with non-zero probability) of infinite dimensional pure states in
the single-copy situation.  In general, entanglement of a pure
state $\ket{a}$ is characterized by the sequence of Schmidt
coefficients $\{ \lambda^a_i \}$ ($0 \leq i \leq d$ and $0 \leq i
\leq \infty$ for finite $d$ and infinite dimensional systems,
respectively).  Recall that the corresponding result for finite
dimensional systems is given by B in Table~\ref{table}; The two
monotones coincide with the Schmidt rank, the number of non-zero
Schmidt coefficients (Vidal's theorem \cite{vidal}).  With a
simple extension of the result obtained for finite dimensional
systems, it is not possible to determine convertibility between
the states with infinite Schmidt ranks.

Since analysis of convertibility between the ``genuine'' infinite
dimensional states (with infinite Schmidt ranks) are our main aim,
we adopt Vidal's theorem to infinite dimensional systems as
following: {\bf Theorem 1 (Vidal)}: $\ket{\psi} \in \Hi$ can be
converted to $\ket{\phi} \in \Hi$ by SLOCC with non-zero
probability in the single-copy situation if and only if there
exists $\epsilon > 0$, $g_\psi (n) / g_\phi (n) \ge \epsilon$ for
all $n \in N$, where $g_a (n) = \sum _{i=n}^{\infty} \lambda^a_i$
is a function defined in terms of Schmidt coefficients $\{
\lambda^a_i \}_{i=0}^\infty$ of a genuine infinite dimensional
state $\ket{a}$.

The function $g_a (n)$ plays the central role in the construction
of the monotones $R^-$ and $R^+$ for infinite dimensional states.
By definition, a sequence of the function $\{g_a (n)\}_{n \in N}$
satisfies four conditions, (strict) positivity $g_a (n)>0$,
(strict) monotonicity $g_a(n) > g_a(n+1)$, convexity $g_a (n+1)
\le \{g_a(n) + g_a (n+2)\}/2$, and normalization $g_a (0)=1$.
Conversely, for a given sequence of functions $\{ g(n)
\}_{n=0}^{\infty}$, there exist a genuine infinite dimensional
state $\ket{a}$, where the Schmidt coefficients are give by
$\lambda^a_i =g(n) -g(n+1)$ if and only if $\{g(n)\}_{n \in N}$
satisfies the strict positivity, strict monotonicity, convexity
and normalization conditions.

According to Theorem 1, if a real parameterized subset $\{
\ket{\xi _r} \} _{r \in A } \subset \Hi$ is totally ordered,
$g_{\xi_r}(n)$ must satisfies $\underline{\lim} _{n \rightarrow
\infty} (g_{\xi_{r_1}} (n) / g_{\xi_{r_2}}(n)) > 0$ if and only if
$r_1 \le r_2$ for all $r_1$ and $r_2$. From the property of
$g_{\xi_r}(n)$, we can construct the monotones $R^-$ and $R^+$
\begin{eqnarray}
    R^-(\psi) &=& \inf \{ r \in A | \underline{\lim} _{n
    \rightarrow \infty}
    g_\psi  (n)/ g_{\xi_r}(n) = 0 \}
    \label{R^-} \\
    R^+(\psi) &=& \inf \{ r \in A | \overline{\lim} _{n
    \rightarrow \infty}
    g_\psi  (n)/ g_{\xi_r}(n) < + \infty
    \}. \label{R^+}
\end{eqnarray}
for all $\{ g_{\xi_r}(n) \} _{r \in A, n \in N}$ satisfying the
three conditions: I. Strict monotonicity for all $r \in A$,  II.
Convexity for all $r \in A$ and $n \in N$, and III.
$\underline{\lim} _{n \rightarrow \infty} (g_{\xi_{r_1}(n)} /
g_{\xi_{r_2}(n)}) > 0$ is equivalent to $r_1 \ge r_2$. The proof
is given by the following:  If $\{ g_{\xi_r}(n) \} _{r \in A, n
\in N}$ satisfies the conditions I, II, and III, the corresponding
set of states $\{ \ket{\xi_r} \} _{r \in A}$ for $\{ g_{\xi_r}(n)
\} _{r \in A, n \in N}$ exists and is totally ordered. Since $A
\in \mathbb{R}$ is assumed to be an interval, the two functions
$R^-(\psi)$ and $R^+(\psi)$ defined by Eqs.~(\ref{rminus}) and
(\ref{rplus}), can be represented by Eqs.~(\ref{R^-}) and
(\ref{R^+}) by using Theorem 1.

%%%%%%%%%%%%%%%%%%%%%%%%%%%%%    Proof    %%%%%%%%%%%%%%%%%%%%%%%%%%%%%%%%%

Now we show that there exists a pairs of genuine infinite
dimensional states which are incomparable to each other. We prove
that the two monotones $R^-(\ket\psi)$ and $R^+(\psi)$ given by
Eqs.~(\ref{R^-}) and (\ref{R^+}) do not necessarily coincide with
each other for infinite dimensional systems, by constructing an
example.

We consider a twice continuously differentiable function $g(x)$,
which is the continuous counterpart of $g(n)$, since a continuous
function is more convenient for analytical investigation.  The
conditions for $g(x)$ to relate to a genuine infinite dimensional
state is now given by $g(x)>0$ (strict positivity), $g^{'}(x) <0$
(strict monotonicity), $g^{''} (x) \ge 0$ (convexity), and $g(0)
=1$ (normalization), for all $x$. If $g(x)$ satisfies the above
conditions except the normalization condition, we can easily
normalize $g(x)$.  Thus we omit the normalization condition for
simplicity.  Since convertibility is determined only by the ratio
of functions, we introduce another function $d(x)$, which is given
by $d(x)=p(x)g(x)$.  Let $d(x)$ satisfy the same conditions as
$g(x)$. Then $p(x)> 0$, $f^{'}(x) p(x) + f(x)p^{'} (x)<0$, $f^{''}
(x) p(x) + 2f^{'}(x)p^{'}(x)+f(x) p^{''}(x) \ge 0$ and $p(1) = 1$
are to be satisfied.

Now we set our function to be $g(x)= e^{-x}$.  Our choice of
$g(x)$ represents one of the most tractable genuine infinite
dimensional entangled states, the two mode squeezed state
$\ket{\psi_q} =\frac{1}{c_q} \sum _{n=0}^{\infty} q^n \ket{n}
\otimes \ket{n}$, where $q$ is a squeezing parameter. We give a
construction of a function $d(x)$ which indicates the existence of
incomparable genuine infinite dimensional states. In this case,
the conditions for $p(x)$ become simple, $p(x) > 0 $, $p(x) -
p^{'}(x)>0$, and $p(x) -2p^{'}(x) +p^{''}(x) \ge0$.

We choose $p(x)$ to be parameterized by $r$ as $p(x)=p_{r}(x) =
(\log x)^{r}\{ \sin (\log x)+1\} +({\log x})^{-1}$ where $0 < r <
+\infty$. We define two functions $m_r(x) \equiv p_r(x) -
p_r^{'}(x)$ and $c_r(x)\equiv p_r(x) -2p_r^{'}(x) +p_r^{''} (x)$
for evaluating monotonicity and convexity, respectively. Then we
have
\begin{eqnarray}
    m_r(x) &=&
    \{(\log x )^{1+r} - (1+r)(\log x)^{r} x^{-1} \}
    \{ \sin (\log x) + 1\} \nonumber \\
    & & {} + ({\log x}){-1} + O((\log x)^{1+r}x^{-1})
\\
    c_r(x) &=&
    \{\sin (\log x) +1\} [(\log x)^{1+r} -
    2(1+r)(\log x)^{r} x^{-1} \nonumber \\
    & & {} -
   (1+r)x^{-1}
    \{(\log x)^{r} - r(\log x)^{r - 1}\}] \nonumber \\
    &+& ({\log x})^{-1} + O((\log x)^{1+r}x^{-1})
\end{eqnarray}

For all $0 < r_1 < r_2 < \infty$, there exists $x_{r_1, r_2}
> 0$ such that $m_r(x)>0$ and $c_r(x) \ge 0$ for all $x \ge x_{r_1, r_2}$,
and $r \in [r_1, r_2]$.  That is, the function $p_r(x+x_{r_1,
r_2})$ satisfies the positivity, monotonicity and convexity
conditions. Therefore we can consider a state $\ket{\xi_r}$
represented by the function $d_r(x)=p_r(x+x_{r_1, r_2}) g(x)$. The
ratio of the functions $g(x)/d_r(x)=1/p_r(x+x_{r_1, r_2})$
determines convertibility between the two states $\ket{\psi_q}$
and $\ket{\xi_r}$. To evaluate the ratio, we rewrite
$p_r(x+x_{r_1, r_2})$ in the discrete form: $p_r(n')=p_r(\Delta n+
x_{r_1, r_2})$ where $\Delta=-\log q$.  Then we can easily show
that $\underline{\lim} _{n \rightarrow \infty} 1/p_r (n) = 0$ and
$\overline{\lim} _{n \rightarrow \infty} 1/p_r (n) = \infty$.
Defining $R^-$ and $R^+$ from $\{ \ket{\xi_r} \}_{r \in
(r_1,r_2)}$, we obtain $R^-(\psi) =  r_1$ and $R^+(\psi) = r_2$.
The two states $\ket{\psi}$ and $\ket{\xi_r}$ for all $r \in
[r_1,r_2]$ are now shown to be incomparable under SLOCC.

%%%%%%%%%%%%%%%%%%%%%%%%% feasibility %%%%%%%%%%%%%%%%%%%%%%%%%%%%%%%%%%%%

The state which corresponds to the function $d(x)$ in our example
may not be feasible to create with present technology. However we
can show the existence of incomparable states by choosing other
forms of the function for $p(x)$, if $g(x)$ is a function
converging as fast as or faster than exponential functions.  Since
the conditions for incomparable states are not related to the
Schmidt basis, we can choose a Schmidt basis which is easy to
control in experiments. We still have the possibility to find more
feasible incomparable states.

Another point related to feasibility in the laboratories is that
our functions $R^-(\ket{\psi})$ and $R^+(\ket{\psi})$ are
discontinuous for the usual topology of Hilbert space. This
discontinuity is caused by the discontinuity of the SLOCC
convertibility itself. Since we cannot completely determine
Schmidt coefficients of the states in realistic situations, we
cannot apply our discussion immediately to such situations.
However, we can say that the maximum probability to convert $\ket{\psi
'}$ where $\| \ket{\psi} - \ket{\psi '} \| < \epsilon $ for small
$\epsilon $ to $\ket{\phi}$ and the probability of the inverse
process are both very small, if $\ket{\psi}$ and $\ket{\phi}$ are
incomparable under SLOCC, because the maximum probability of
convertibility under SLOCC itself is continuous. In other words,
this incomparability-like property appears in the limit of large
dimensional space.

%%%%%%%%%%%%%%%%%%%%%% summary %%%%%%%%%%%%%%%%%%%%%%%%%%%%%%%

In this paper, we have developed a general formulation for
constructing a pair of convertibility monotones using order
properties.   The monotones are considered as generalizations of
distillable entanglement and entanglement cost. This formulation
can be applied to many different situations to analyze
entanglement convertibility.  We have applied the formulation to
SLOCC convertibility for genuine infinite dimensional pure states
in the single-copy situation.  By constructing an example, we have
proved the existence of SLOCC incomparable pure bipartite states,
a new property of entanglement in infinite dimensional systems. In
contrast, incomparable pure states only exists for multipartite
systems (such as GHZ and W states for three qubit states) in
finite dimensional systems.

One of the important remarks in this letter is that the ordering
property under SLOCC convertibility is changed fundamentally, from
total ordering to non-total (partial) ordering, with the shift in
dimensionality from finite to infinite. It had been widely
believed that the fundamental entanglement properties of finite
and infinite dimensional systems are similar. However, we have
shown that there exists a significant difference in
convertibility. Our result encourages the search for other
fundamental differences between finite or infinite dimensional
systems.

%%%%%%%%%%%%%%%%%%%%%% acknowlegement %%%%%%%%%%%%%%%%%%%%%%%%%%%%%%%

The authors are grateful to M. Hayashi for helpful mathematical
comments. This work is supported by the Sumitomo Foundation, the
Asahi Glass Foundation, and the Japan Society of the Promotion of
Science.

%%%%%%%%%%%%%%%%%%%%%%%%%%bibliography%%%%%%%%%%%%%%%%%%%%%%%%%%%%%%%%%

\end{document}